\documentclass[preprint,aps]{revtex4}

\usepackage{graphicx}
\usepackage{dcolumn}
\usepackage{bm}
\usepackage{epsfig}

\begin{document}
\newcommand{\gsim}{\hbox{\rlap{$^>$}$_\sim$}}
\newcommand{\lsim}{\hbox{\rlap{$^<$}$_\sim$}}

\title{Fast Extragalactic X-ray Transients \\
        From Binary Neutron Star Mergers }

\author{Shlomo Dado and Arnon Dar}
\affiliation{Physics Department, Technion, Haifa, Israel}

\begin{abstract}
The observed light curves and other properties of the two extragalactic 
fast x-ray transients, CDF-S XT1 and CDF-S XT2, which were discovered 
recently in archival data of the Chandra Deep Field-South (CDF-S) 
observations, indicate that they belong to two different populations of 
X-ray transients. XT1 seems to be an x-ray flash (XRF), i.e., a 
narrowly beamed long duration gamma ray burst  viewed from far 
off-axis while  XT2 seems to be a nebular emission powered by a 
newly born millisecond pulsar in a neutron stars binary merger. 

\end{abstract}


\maketitle

\section{Introduction}.

The production of gamma ray bursts (GRBs) in neutron stars binary (NSB) 
mergers due to gravitational wave (GW) emission was first suggested in 1984 
[1] to be due to explosion (later called a kilonova [2]) of the lighter 
neutron star in NSB mergers following tidal mass loss. Later in 1987, it was 
suggested that GRBs are produced by neutrino annihilation fireballs [3] 
formed around the newly born neutron star remnant in NSB mergers driven by 
gravitational wave (GW) emission and in stripped envelope supernova 
explosions. But, the observations of GRBs by the Compton Gamma-Ray 
Observatory (CGRO), soon after its launch in 1991, indicated that neither one 
of these two mechanisms was powerful enough to produce observable GRBs at the 
very large cosmological distances [4] indicated by the CGRO observations. 
Thus, in 1994, the fireball mechanism [5] for production of prompt emission 
in GRBs was replaced by inverse Compton scattering (ICS) of external light by 
a highly relativistic bipolar jets of plasmoids of ordinary plasma, launched 
by fall back ejecta on the remnants of NSB mergers and stripped envelope 
supernova explosions and in the collapse of neutron stars to quark stars or 
black holes due to mass accretion in high mass X-ray binaries [6]. Shortly 
after the discovery in 1997 by the Italian-Dutch BeppoSAX satellite [7] discovered 
that GRBs are followed by a long-lived x-ray afterglow, and the following 
discoveries of their afterglows at longer wave lengths [8], the jet model 
of GRBs [6] was extended to predict also the properties of GRB afterglows 
[9]. The spherical fireball model, however, which predicted the existence 
of GRB  afterglows with a single power light curves [10], was later 
replaced, 
first by conical $e^+e^-\gamma$ fireballs [11], and finally by conical jets 
of ordinary plasma [12].

The first NSB merger detected in gravitational waves by the Ligo-Virgo 
detectors [13], GW170817, was followed $1.74\!\pm\!0.5$ s later by a short 
duration gamma ray burst GRB170817A [14]. Moreover, very large base 
interferometry (VLBI) late time follow up observations of the radio afterglow 
of GRB170817A provided the first direct observational evidence [15] that NSB 
mergers launch narrowly collimated jets of "superluminal" plasmoids which 
produce narrowly beamed short gamma ray burst (SGRB) pulses with narrowly 
beamed late time afterglow [15]. However, its early-time optical afterglow 
was claimed to be produced by a kilonova [16], or a nebular emission with a 
thermalized bremsstrahlung, powered by a millisecond pulsar (MSP) remnant of 
the NSB merger [17]. The claimed kilonova, however, had an early time 
bolometric light curve much different from that claimed before to be 
associated with SGRB [18], while its bolometric light curve displayed the 
same "universal shape" [19] as that of about half of the well sampled early time 
x-ray afterglows of ordinary (viewed near axis) SGRBs, which were measured 
before with the Swift XRT [20].

GW170817/GRB170817A was the first indisputable evidence that NSB mergers 
produce SGRBs with a narrowly beamed late time afterglow, and perhaps an 
early time isotropic afterglow emission. By chance, two days before 
this event, it was predicted [20] that the visible smoking guns of NSB 
mergers are mostly millisecond pulsars remnants whose spin down powers an
isotropic nebular emission with nearly a universal shape light curve [21].

Recently, two extragalactic fast x-ray transients, CDF-S XT1 [22] associated 
with a faint galaxy at a photometric redshift $z_{ph}\!\sim\!2.2$ and CDF-S 
XT2 associated with a galaxy at redshift $z\!=\!0.738$ [23] were discovered 
in archival data of the Chandra Deep Field-South (CDF-S) observations. It was 
suggested that both are X-ray nebular emission powered by the spin down of 
newly born magnetars in NSB mergers, which produced 
SGRBs that are beamed away from Earth [23,24]. In 
this paper, we provide supportive evidence that CDF-S XT2 was an early time 
isotropic orphan x-ray afterglow of NSB merger which produced a remnant 
MSP. However, the light curve of CDF-XT1 indicates that, 
more likely, it was an X-ray flash (XRF) i.e., a prompt emission pulse of a 
narrowly beamed long duration gamma ray burst (LGRB) viewed from far off axis 
[38], rather than a nebular emission powered by the spin down of a newly born 
MSP [26].

\section{Orphan x-ray afterglows of GRBs}

There is mounting observational data indicating that LGRBs and SGRBs are 
narrowly beamed along the direction of motion of highly relativistic jets 
of plasmoids which produce them [27]. Distant observers located outside 
the beaming cone of such jets miss their prompt gamma ray emission and 
their beamed afterglows. However, the deceleration of jets in the 
circumburst medium decreases their bulk motion Lorenz factor $\gamma(t)$ 
and widens their beaming cone of their radiation. Once their beaming cone 
includes the distant observer, their afterglow radiations become visible. 
However, so far, no such orphan GRB afterglows [28] have been detected. 
That could have been because of various reasons, such as lack of a unique 
signature, a luminosity below detection threshold by the time their 
beaming cone has expanded enough to include Earth, a very small sky 
coverage in very deep searches, and a small signal to background ratio.

Another type of orphan electromagnetic afterglows, which can be seen from 
large cosmic distances, is an isotropic early time nebular emission powered 
by a newly born MSP [21]. Such MSPs can be the remnants of NSB mergers which 
do not not produce a GRB, or which produce narrowly beamed GRBs that do not 
point to Earth. In both cases they can produce an x-ray emission 
with a "universal" shape [19]. Indeed, a large fraction of SGRBs which point to 
Earth and have a well sampled early time x-ray afterglow light curve have a 
universal shape expected from a nebular emission powered by a newly born MSP 
[19]. Such afterglows can be seen up to very large cosmic distances. 
Moreover, if the birth of an MSP is accompanied by a narrowly beamed SGRB 
which points away from Earth, then the x-ray nebular emission appears as an 
orphan extragalactic fast x-ray transient [21].

Below, we show first that the extragalactic fast X-ray transient CDF-S XT2 
discovered in the Chandra Deep Field-South observations [22,23] and a large 
fraction of the well sampled early time X-ray afterglow of SGRBs [19] share 
the same "universal shape" light curves [19] powered by newly born MSPs [29].
Next we show that the estimated full sky rate of CDF-S XT2 
like events [24] is consistent with that expected from the local cosmic rate 
of neutron star mergers [30]. Together, they provide supporting evidence to the 
suggestion that CDF-S XT2 [23] probably is the early time nebular 
emission powered by the spin down of a newly born pulsar of 
a far off axis SGRB  produced in NSB merger [21,23,24,25].

\section{Nebular emission powered by newly born MSPs}
The spin down energy of a pulsar with a constant 
moment of inertia $I$, is given by
\begin{equation}
\dot{E}=4\, \pi^2\,\nu\,\dot{\nu}\, I 
\end{equation} 
where $\nu$ is its spin frequency. 
For a pulsar with braking index  $n$ defined by
\begin{equation}
\dot{\nu}=-k\,\nu^n\,
\end{equation}
where $k$ is a time independent constant,
the  rate of its  rotational energy loss 
is given by 
\begin{equation}
\dot{E}(t)=\dot{E}(0)(1+t/t_b)^{-(n+1)/(n-1)},
\end{equation}
with
\begin{equation}
t_b=-\nu(0)/(n-1)\dot{\nu}(0)=P(0)/(n-1)\dot{P}(0),
\end{equation}
where $P=1/\nu$ is the pulsar's period.

For a spin down dominated by the emission of magnetic dipole 
radiation (MDR) in vacuum  $n\!=\!3$ and
\begin{equation}
L(t)=L(0)/(1+t/t_b)^2,
\end{equation}
where $L(t)=\dot{E}$.
As long as the early time x-ray afterglows of SGRBs from 
NSB mergers are pulsar wind nebula (PWN) emission powered by a 
constant fraction $\eta$ of the spin down energy of a
newly born pulsar with a braking index $n=3$,
the early time x-ray 
afterglow of both a visible and an invisible SGRBs have the  
universal behavior,
\begin{equation}
F_x(t)/F_x(0)=[1+t_s]^{-2}
\end{equation} 
where $F_x(t)$ is the measured energy flux of the X-ray 
afterglow of the SGRB and $t_s=t/t_b$.    

In Figure 1 the reported x-ray light curve of CDF-S XT2  
[23], reduced to the dimensionless universal form given by 
Eq.(6), is compared to the early time light curves of the 
x-ray afterglow of about half of the SGRBs with a well 
sampled x-ray afterglow measured with the Swift XRT and reported in  
the Swift-XRT Light Curves Repository [20].
For each SGRB afterglow 
the values of the parameters $F_x(0)$ and $t_b$ 
were obtained from a best fit of Eq.(6) to the the measured 
light curves. 
Their values were reported  in Table I of [19]. 
A best fit of Eq.(6) to the  0.3-10 keV 
X-ray light curve of CDF-S XT2 [23]
has yielded the best fit values,
$F_x(0)=8.8\times 10^{-13}\, {\rm erg/cm^2\, s}$
and $t_b=1705$ s.

\begin{figure}[]
\centering
\epsfig{file=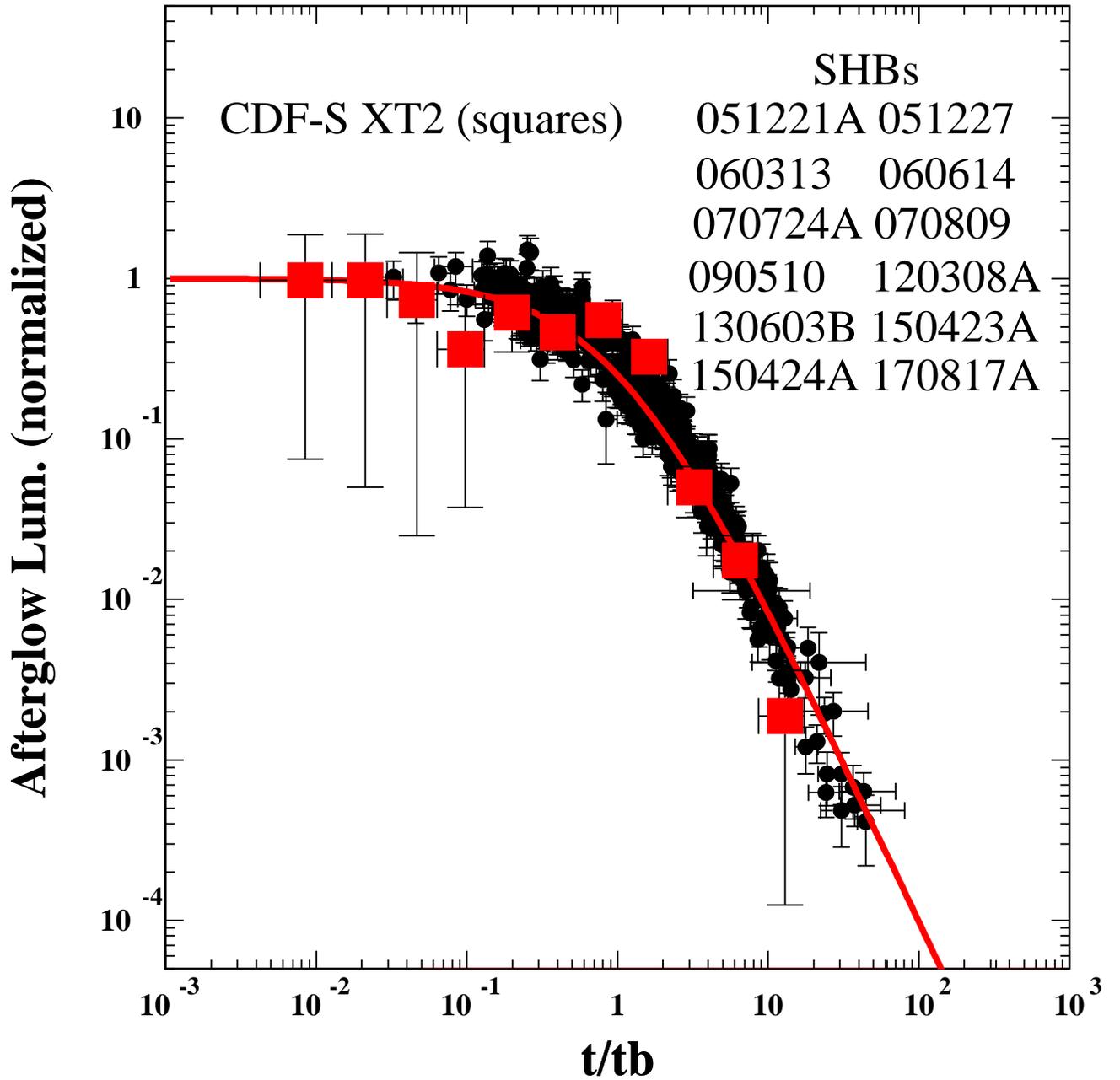,width=19cm,height=19cm}
\caption{Comparison between the scaled 0.3-10 keV
light curves of the well sampled X-ray afterglow 
of SGRBs during the first couple of days after 
burst measured with the Swift XRT [20] 
and the 0.3-10 keV light curve of CDF-S XT2 [23]. 
The line is the expected  universal behavior 
given by Eq.(6) of a PWN afterglow powered by a 
newly born millisecond pulsar with a braking index $n=3$.} 
\end{figure}

\section{Lower bound on MSP magnetic field}
If the spin down of the newly born pulsar is dominated
by magnetic dipole radiation, the magnetic field $B_p$ 
at the pulsar's magnetic poles satisfies [31]
\begin{equation}
B\,sin\alpha\approx 6.8\times 10^{19}\,[P\,\dot{P}]^{1/2}\,\, Gauss, 
\end{equation}
where $P$ is in seconds and $\alpha$ is the angle of the 
magnetic poles relative to the rotation axis.  

The initial period of the pulsar could be estimated [19] from the best 
fit parameters $F_x(0)$, $t_b$ and the luminosity distance of the PWN 
only when the fraction $\eta$ of entire spin down energy of the pulsar, 
which has been converted by the PWN to the observed afterglow of the 
SGRB, is known. However, usually the exact geometry of the PWN and the 
fraction of the pulsar spin energy converted to X-ray emission in the 
PWN are not known. As a result the value of $\eta$ is usually unknown. 
Moreover there is no reliable evidence that millisecond pulsars spin 
down by the emission of magnetic dipole radiation. That, and the lack 
of reliable evidence that MSPs spin down by magnetic dipole radiation 
[32] prevents the use of Eq.(7) to obtain a reliable estimate of the 
magnetic field of the neutron star at the magnetic poles.

However, if the widespread assumption that MSPs spin down mainly by 
magnetic dipole radiation is correct, then a lower bound on the 
magnetic field at the poles can be estimated from the best fit value of 
$t_b$ obtained from the best fit of Eq.(6) to the early X-ray afterglow 
of SGRBs powered by newly born pulsars, as follows. Substitution of the
lower classical limit $P\geq 2\,\pi\, R/c\approx 0.2$ ms for a canonical 
pulsar with a radius $R\approx 10$ km and a surface velocity equal to 
the speed of light, and substituting it in Eq. (6), and the use of 
the relation
$\dot{P}(0)\!=\!P(0)/2\,t_b$ valid for a braking index $n=3$, which is 
valid for a constant magnetic field in vacuum, imply the lower limit,
\begin{equation}
B_p(0)\,\gsim\, 10^{16}\,\sqrt{(1+z)/(t_b/s)}\,\,\, Gauss.
\end{equation}
Eq.(8) yields $B_p(0)\,\gsim\, 3\times 10^{14}$ Gauss for CDF-S XT2
at its redshift $z=0.735$ [23]. 

The best fit of eq.(6) to the early time optical afterglow [33] of 
GRB170817A at $z\!=\!0.0093$, has yielded $t_b\!\approx\! 117374$ s, 
[19]. Hence, eq.(8) yields $B_p(0)\,\gsim\,10^{14}$ Gauss for a
pulsar powering the early time  bolometric light curve of the 
optical afterglow of GRB170817A. 

\section{The full sky rate of orphan MSP powered afterglows}

If the cosmic rate of NSB mergers in a comoving volume  as a function 
of redshift $z$ is proportional to the star formation rate, SFR(z), 
e.g., if they are produced mainly by fission of a fast rotating 
core in core collapse supernova explosions of massive stars 
[17],  then 
the production rate of pulsar powered afterglows by NSB mergers
in a  comoving volume is given by [34]
\begin{equation}
{d\dot {N}\over dz}\propto SFR(z){dV_c(z)\over dz}\,{1\over 1+z}\,
\end{equation}
where $V_c(z)$ is the comoving volume at redshift $z$.
In a standard $\Lambda$CDM cosmology, $dV_c(z)/dz$ is given by
\begin{equation}
{dV_c(z)\over dz}={c\over H_0}\, {4\,\pi\,[D_c(z)]^2 \over 
                   \sqrt{(1+z)^3\Omega_M +\Omega_\Lambda}}\,,
\end{equation}
where $H_0$ is the current Hubble constant, $\Omega_M$ and 
$\Omega_\Lambda$ are, respectively, the current density of ordinary energy 
and of dark energy, in critical energy-density units, and $D_c(z)$ is
the comoving distance at a red shift $z$, which satisfies
\begin{equation}
D_c(z)= {c\over H_0}\, \int_0^z {dz'\over \sqrt{(1+z')^3\Omega_M 
+\Omega_\Lambda}}\,.
\end{equation}

In order to estimate the full sky rate of NSB mergers, $\dot N(z)$, 
as given by Eqs.(9-11) we have adopted the current best values of the 
cosmological parameters obtained from the combined WMAP and Planck 
data [35]:  a Hubble constant $H_0=67.4\,{\rm km/s\, Mpc} $, 
$\Omega_M=0.315$ and $\Omega_\Lambda= 0.685$ and the SFR(z) compiled 
and standardized in [36] and [37] from optical measurements. This 
standardized SFR(z) is well approximated [33] by a log-normal 
distribution,
\begin{equation}
{\rm SFR}(z)\approx 0.25\, e^{-[ln((1+z)/3.16)]^2/0.524}\,\,
                 {\rm M_\odot\, Mpc^{-3}\, y^{-1}}\,. 
\end{equation} 

Assuming that the cosmic rate of neutron star mergers (NSMs) as a
function of redshift is proportional to SFR(z) 
given by Eq.(12), and that the rate of NSMs in a comoving volume of 
Gpc$^3$ is $ (1540\, +3200/-1200)\, Gpc^{-3}\, y^{-1}$, as estimated in [13] 
from the Ligo-Virgo GW observations, then the expected full sky rate 
$\dot{N}(\leq z)$ of NSMs in the standard cosmological model with the 
updated values of the cosmological parameters measured with Planck [34], 
is shown in Figure 2. This rate, to a good approximation, is also the 
expected rate of orphan early time afterglows produced by the majority of 
SGRBs which point away from Earth. Their full sky rate obtained 
from their estimated rate $59+77/-38$ evt ${\rm y^{-1}\, deg^{-2}} $
in [24] from the CDF-S observations of XT1 and XT2 [24], after 
subtracting the contribution of XT1, is also indicated in 
Figure 2.
\begin{figure}[]
\centering
\epsfig{file=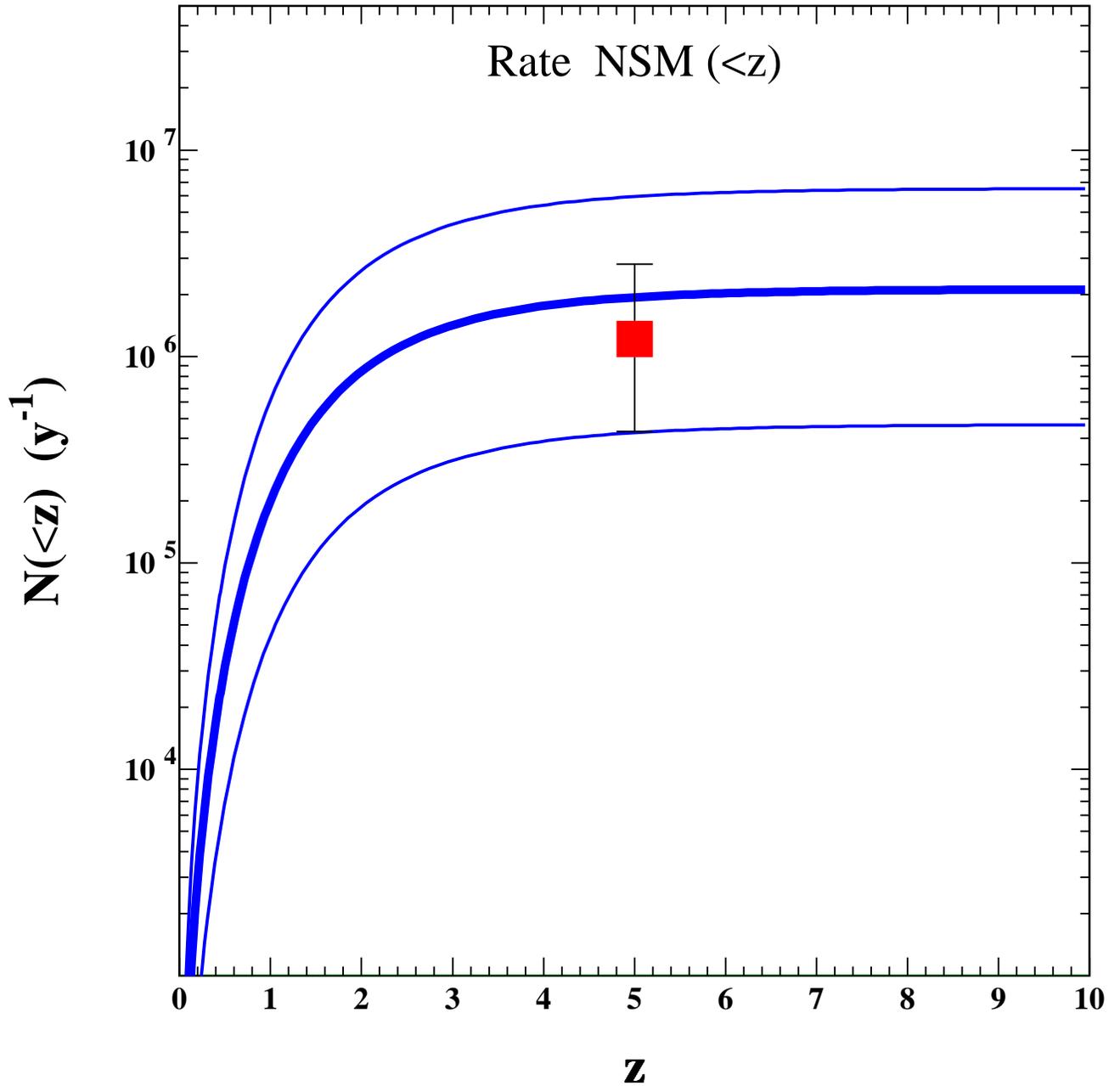,width=19cm,height=19cm}
\caption{The expected full sky rate  of neutron stars  merger (NSM) 
with redshift $\leq z$, as a function of $z$. The calculated rate is  
based on the standard cosmological model and the
assumption that the NSM rate  as a function of redshift 
$z$ is proportional to the observed star formation rate, SFR(z), 
as parametrized in eq.(12). The full and thin lines correspond to 
the estimated rate and its errors in a comoving $Gpc^3$ volume reported
in [13] by the Ligo-Virgo collaboration.
The inserted point is the full sky rate   
estimated in [24], after subtracting the contribution of  CDF-S XT1.}
\end{figure}

\section{CDF-S XT1 as far off-axis LGRB}

In the cannonball model of GRBs, the predicted pulse shape above a minimal 
energy $E_m$ of prompt emission pulses of GRBs has the behavior [26,38],
\begin{equation}
{dN_\gamma(E>E_m)\over dt}\!\propto\!{t^2\, exp[-E_m/E_p(t)]
                       \over (t^2\!+\!\Delta^2)^2}\,,
\end{equation}
where $E_p(t)$ is the peak energy at time $t$.
For XRFs, i.e., far off axis GRBS, the 
exponential factor on 
the right hand side of eq.(13) can be neglected, which yields a 
full width at half maximum ${\rm FWHM}\!\approx 2\,\Delta$, a rise 
time $RT\!\approx \!0.59\,\Delta$ from half peak to peak  at 
$t=\Delta$, and a decay time $DT\!\approx \!1.41\,\Delta$ from 
peak to half peak. This pulse shape is in good agreement 
with that observed in resolved, well sampled, GRB pulses [17,27].
Figure 3 shows a comparison between eq.(13) and  the 
measured light curve of CDF-S XT1. The best fit 
normalization, a pulse start up time $t_0\!=\!38.8$ s, and  
$\Delta\!=\!50.6$ s, yield $\chi^2/dof\!=\!4.52/9\!=\!0.50$.
\begin{figure}[]
\centering
\epsfig{file=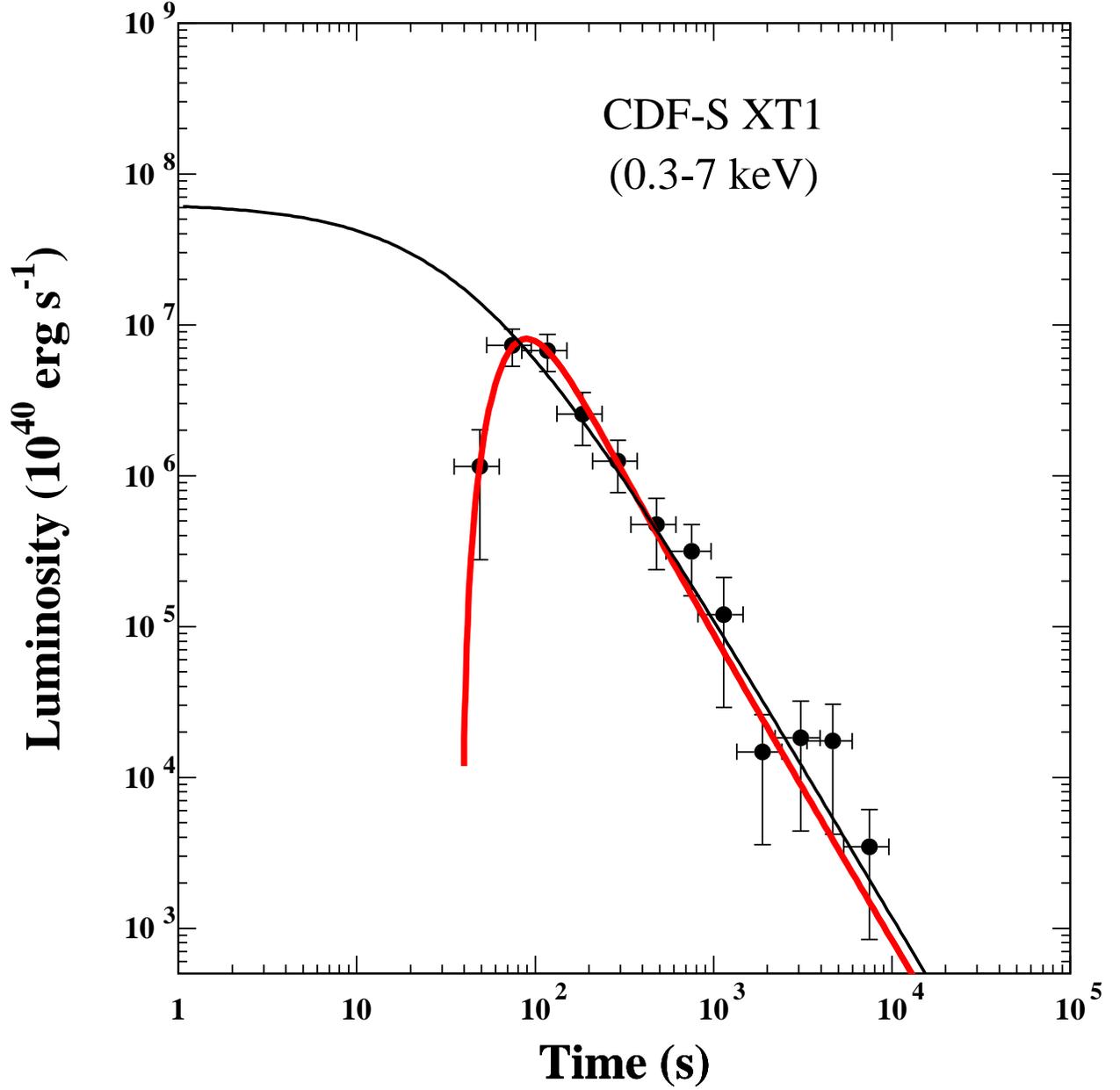,width=18.cm,height=18.cm}
\caption{Comparison between the observed
pulse shape of CDF-S XT1 [22]  in the 0.3-7 keV
x-ray band and (a) the expected pulse shape of far off axis LGRB pulse 
as given by eq.(13) (thick red line) with the best 
fit parameters listed in the text,  which  yield $\chi^2/df=0.50$,
(b) the best fit universal shape of a nebular x-ray emission 
powered by the spin down of a newly born MSP (thin black line),
as given by eq.(5).}
\end{figure}
\section{Conclusions}
The observed light curve of CDF-S-XT2 recently discovered with the 
Chandra x-ray observatory [23] and the estimated full sky rate of such 
extragalactic fast x-ray transients support the conclusion 
that it was an early time nebular emission powered by a newly born 
millisecond pulsar with a strong magnetic field, $B\,\gsim\,10^{14}$ 
Gauss in a neutron stars binary merger which produced SGRB beamed away 
from the direction of Earth [21,23]. The estimated strength of the 
dipole magnetic field of these newly born pulsars, from the afterglow 
which they power, depends on the {\it assumption} that their spin down 
is dominated by magnetic dipole radiation, which may or may not be 
true. The typical signature of orphan afterglows of SGRBs - a fast 
rise after burst followed by a short plateau phase of typically a few 
thousands of seconds, which turns into a fast temporal decline, may 
partially explain why such transients have not been found so far in 
searches of electromagnetic afterglows of SGRBs from the nearby binary 
neutron star merger candidates detected in gravitational wave by 
Ligo-Virgo in run O3 [39]. Fast extragalactic x-ray transients, such 
as CDF-S XT1 [22], which unlike CDF-S XT2 [23], do not have the "universal 
shape" of an early time pulsar powered afterglow of SGRB [19]. Its  
light curve has the universal shape of well resolved 
prompt emission pulses of LGRBs viewed far off axis, i.e. of 
X-ray flashes (XRFs), which were discovered in archival CDF data [40]
long before CDF-S XT1 [22].

{\bf Acknowledgenents}: We thank P. Jonker, Y. Q. Xue and an anonymous   
referee for useful comments.

\end{document}